\documentclass[12pt]{article}
\usepackage{epsf}
\usepackage{graphicx}

\title{ {\bf QCD Sum Rules Study of the Semileptonic $B_{s}(B^{\pm})(B^{0})\rightarrow D_{s}[1968](D^{0})(D^{\pm}) l\nu $ Decays }}
\author{\vspace{1cm}\\
K. Azizi \thanks {e-mail: e146342@metu.edu.tr} \\ Physics
Department, Middle East Technical University, 06531\\ Ankara, Turkey
}
 \date{}
\begin{document}
\setlength{\baselineskip}{24pt} \maketitle
\setlength{\baselineskip}{7mm}
\begin{abstract}
The form factors of the semileptonic $B_{q}\rightarrow
D_{q}(J^{P}=0^{-})\ell\nu$ with $q=s, u, d$ transitions are
calculated in the framework of three point QCD sum rules. Using the
$q^2$ dependencies of the relevant form factors, the total decay
width and the branching ratio for these decays are also evaluated. A
comparison of our results for the form factors of $B\rightarrow
D\ell\nu$ with the lattice QCD predictions  within heavy quark
effective theory and zero recoil limit is presented. Our results of
the branching ratio are in good agreement with the constituent quark
meson  model for ($q=s, u, d$) and experiment for ($q= u, d$). The
result of branching ratio for $B_{s}\rightarrow D_{s}(1968)\ell\nu$
indicates that this transition can also be detected at LHC in the
near future.
\end{abstract}
\thispagestyle{empty}
\newpage
\setcounter{page}{1}
\section{Introduction}

The pseudoscalar $B_{q}$ meson decays are very promising tools to
constrain the Standard Model (SM) parameters, explore heavy quark
dynamics and  search for new physics. The semileptonic decays of
heavy flavored mesons  are also useful for determination of the
elements of the Cabibbo-Kobayashi-Maskawa (CKM) matrix, leptonic
decay constants as well as the origin of the CP violation. Neutral
$B_{s}^{0}$ and $B_{d}^{0}$ meson decays are  interesting  to study
CP violation.

When LHC begins operation, an abundant number of $B_{q}$ mesons will
be produced. This will provide a real possibility for studying the
properties of the $B_{q}$ mesons and their various decay channels.
Some possible decay channels  of $B_{q}$ mesons are  their
semileptonic decays to $D_{q}l\nu$ . The $B_{q} \rightarrow D_{q}
\ell\nu$ transitions occur via the $b\rightarrow c$ transition with
s, d or u as spectator quarks. The most common decay mode of B
mesons is clearly $b \rightarrow c$ transition, since it is the most
dominant transition among the b quark decays. The semileptonic
$B_{q} \rightarrow D_{q} \ell\nu$ decays are interesting because
they could play a fundamental role in probing new physics charged
Higgs contributions in low energy observables. Moreover, they open a
window onto the strong interactions of the constituent quarks of the
pseudoscalar $D_{s}$ meson and could give useful information about
the structure of this meson (for a discussion about the nature of
$D_{sJ}$ mesons and their quark content see \cite{Colangelo1,
Swanson}). Analysis of the $D_{s_{0}}(2317)\rightarrow
D_{s}^{\ast}\gamma$, $D_{sJ}(2460)\rightarrow D_{s}^{\ast}\gamma$
and $ D_{sJ}(2460)\rightarrow D_{s_{0}}(2317)\gamma$ indicates that
the quark content of these mesons is probably $\overline{c}s$
\cite{colangelo2}.

 The long distance dynamics of such type transitions are parameterized in
terms of some form factors, which are related to the structure of
the initial and final meson states. For calculation of these form
factors which play fundamental role in the analysis of these
transitions, some nonperturbative approaches are needed. Among the
existing nonperturbative methods, QCD sum rules has  received
especial attention, because this approach is based on the
fundamental QCD Lagrangian. There are two kinds of QCD sum rule
approaches, three point and light cone QCD. In three point QCD sum
rules, the perturbative part of the correlation function is expanded
in terms of operators having different mass dimensions with the help
of the operator product expansion (OPE). In light cone QCD, the
distribution amplitudes (DA's) of the particles expanding in terms
of different twists are used \cite{braun, aliev1, aliev2}. This
method has been applied successfully for wide variety of problems
\cite{ kazem1, aliev3, aliev4, aliev5, aliev6} (for a review see
also \cite{colangelo3}). In present work, we describe the
semileptonic $B_{q} \rightarrow D_{q} \ell\nu$ decays by calculating
the relevant form factors in the framework of the three point QCD
sum rules approach. Note that, the form factors of $ B \rightarrow D
\ell\nu$ have been calculated in lattice QCD \cite {Hashimoto,
Okamoto, Divitiis1, Divitiis2} and  the subleading Isgur-Wise form
factor is computed in QCD sum rules and its application for the $B
\rightarrow D \ell\nu$ decay is shown in
 \cite {Neubert, Ligeti}(for similar previous works see also \cite{ovch, groz1, groz2} ). Moreover, the
$B_{q} \rightarrow D_{q} \ell\nu$ transitions have been studied in
the constituent quark meson (CQM) model for $q =s, u, d$ in
\cite{zhao} and for $q = u, d$, the experimental results can be
found in \cite{Yao}.

This paper is organized as follows: In section II, we calculate the
sum rules for the two form factors relevant to these transitions.
Section III is devoted with the numerical analysis, conclusion,
discussion and  comparison of our results for the form factors and
branching ratios with those  of the other phenomenological model,
lattice QCD and experiment.
\section{Sum rules for the $ B_{q}\rightarrow D_{q}\ell\nu $ transition form factors}
In the quark level, the $B_{q} \rightarrow  D_{q}\ell\nu $
transitions proceed by the $b\rightarrow c$ transition  (see Fig.
1).
\begin{figure}
\vspace*{-1cm}
\begin{center}
\includegraphics[width=8cm]{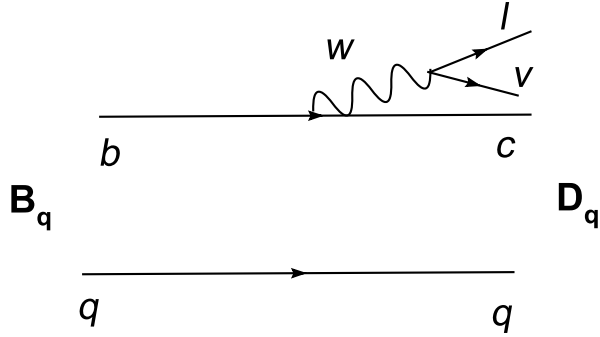}
\end{center}
\caption{$B_{q}\rightarrow D_{q}\ell\nu $ transitions at tree level
} \label{fig1}
\end{figure}
The matrix element for these transitions at the quark level can be
written as:
\begin{equation}\label{lelement}
M_{q}=\frac{G_{F}}{\sqrt{2}} V_{cb}~\overline{\nu}
~\gamma_{\mu}(1-\gamma_{5})l~\overline{c}
~\gamma_{\mu}(1-\gamma_{5}) b .
\end{equation}
To obtain the matrix elements for $B_{q}\rightarrow D_{q}\ell\nu$
decays, we need to sandwich Eq. (\ref{lelement}) between initial and
final meson states, so the amplitude of these decays gets the
following form:
\begin{equation}\label{2au}
M=\frac{G_{F}}{\sqrt{2}} V_{cb}~\overline{\nu}
~\gamma_{\mu}(1-\gamma_{5})l<D_{q}(p')\mid~\overline{c}
~\gamma_{\mu}(1-\gamma_{5}) b\mid B_{q}(p)>.
\end{equation}
Our aim is to calculate  the matrix elements
$<D_{q}(p')\mid\overline{c}\gamma_{\mu}(1-\gamma_{5}) b\mid
B_{q}(p)>$ appearing in Eq. (\ref{2au}). Because of parity and
Lorentz invariance the axial vector part of transition current,
$~\overline{c}~\gamma_{\mu}(1-\gamma_{5}) b~$,  does not have any
contribution to the matrix element considered above, so the
contribution comes only from the vector part of the transition
current. Considering the parity and  Lorentz invariances, one can
parameterize this matrix element  in terms of the form factors in
the following way:
\begin{equation}\label{3au}
<D_{q}(p')\mid\overline{c}\gamma_{\mu} b\mid
B_q(p)>=f_{1}(q^{2})P_{\mu}+f_{2}(q^{2})q_{\mu},
\end{equation}
where $f_{1}(q^2)$, $f_{2}(q^2)$ are the transition form factors and
$P_{\mu}=(p+p')_{\mu}$, $q_{\mu}=(p-p')_{\mu}$.

From the general philosophy of QCD sum rules method, in order to
calculate the form factors we consider  the following correlator:
\begin{eqnarray}\label{6au}
\Pi _{\mu}(p^2,p'^2,q^2)=i^2\int d^{4}xd^4ye^{-ipx}e^{ip'y}<0\mid
T[J _{D_{q}}(y) J_{\mu}^{t}(0) J_{B_{q}}(x)]\mid  0>,
\end{eqnarray}
where $J _{D_{q}}(y)=\overline{c}\gamma_{5}q$ and
$J_{B_{q}}(x)=\overline{b}\gamma_{5}q$ are the interpolating
currents of the $D_{q}$ and $B_{q}$, respectively and
$J_{\mu}^{t}(0)=~\overline{c}\gamma_{\mu}b $ is the transition
current.

To calculate the phenomenological or physical  part of the
correlator given in Eq. (\ref{6au}), two complete sets of
intermediate states with the same quantum numbers as the currents
$J_{D_{q}}$ and $J_{B_{q}}$ respectively are inserted. As a result
of this procedure, we get the following representation of the
above-mentioned correlator:
\begin{eqnarray} \label{7au}
\Pi _{\mu}&=&\frac{<0\mid J_{D_{q}}(0) \mid
D_{q}(p')><D_{q}(p')\mid J_{\mu}^{t}(0)\mid B_{q}(p)><B_{q}(p)\mid
J_{q}(0)\mid 0>}{(p'^2-m_{D_{q}}^2)(p^2-m_{B_{_{q}}}^2)}\nonumber \\
&+& \cdots,
\end{eqnarray}
 where $\cdots$ represents the contributions coming from higher states and continuum. The following matrix
 elements in Eq. (\ref{7au}) are defined in the standard way as:
\begin{eqnarray}\label{8au}
<0\mid J_{D_{q}} \mid D_{q}(p')>=-i \frac{f_{D_{q}}m_{D_{q}}^{2}}{m_{c}+m_{q}}, \nonumber \\
<B_{q}(p)\mid J_{B_{q}}\mid
0>=-i\frac{f_{B_{q}}m_{B_{q}}^2}{m_{b}+m_{q}},
\end{eqnarray}
where $f_{D_{q}}$ and $f_{B_{q}}$  are the leptonic decay constants
of $D_{q}$ and $B_{q}$ mesons, respectively. Using Eq. (\ref{3au})
and Eq. (\ref{8au}), Eq. (\ref{7au}) can be written  in hadronic
language as:
\begin{equation}\label{7au1}
    \Pi_{\mu}(p^2,p'^2,q^2) =
    \Pi_{1}(p^2,p'^2,q^2)P_{\mu}+\Pi_{2}(p^2,p'^2,q^2)q_{\mu},
\end{equation}
Where,
\begin{eqnarray}\label{9amplitude}
\Pi_{1}(p^2,p'^2,q^2)&=&-\frac{1}{(p'^2-m_{D_{q}}^2)(p^2-m_{B_{_{q}}}^2)}\frac{f_{D_{q}}m_{D_{q}}^{2}}{m_{c}+m_{q}}\frac{f_{B_{q}}m_{B_{q}}^2}{m_{b}+m_{q}}f_{1}(q^{2})
\nonumber\\&+& \mbox{excited states,}
\nonumber \\
\Pi_{2}(p^2,p'^2,q^2)&=&-\frac{1}{(p'^2-m_{D_{q}}^2)(p^2-m_{B_{_{q}}}^2)}\frac{f_{D_{q}}m_{D_{q}}^{2}}{m_{c}+m_{q}}\frac{f_{B_{q}}m_{B_{q}}^2}{m_{b}+m_{q}}f_{2}(q^{2})
\nonumber\\&+&\mbox{excited states.}
\end{eqnarray}

 Now, let calculate the theoretical part (QCD side) of the correlation function $\Pi
_{\mu}(p^2,p'^2,q^2)$ in quark and gluon languages
    with the help of the operator product expansion(OPE) in the deep Euclidean
region  $p^2 \ll (m_{b}+m_{q})^2 $ and $p'^2 \ll (m_{c}+m_{q})^2$.
The correlator is written in terms of the perturbative and
nonperturbative parts as:
\begin{eqnarray}\label{7au2}
    \Pi_{\mu}(p^2,p'^2,q^2) &=& \bigg[ \Pi^{per}_{1}(p^2,p'^2,q^2)+\Pi^{non-per}_{1}(p^2,p'^2,q^2)\bigg]P_{\mu}\nonumber\\
    &+&\bigg[\Pi^{per}_{2}(p^2,p'^2,q^2)+\Pi^{non-per}_{2}
    (p^2,p'^2,q^2)\bigg]q_{\mu}.
\end{eqnarray}
To obtain the sum rules for the form factors, the two different
representations of $\Pi _{\mu}(p^2,p'^2,q^2)$ are equated. The
theoretical part of the correlator is calculated by means of OPE,
and up to operators having dimension $d=5$, it is determined by the
bare-loop (Fig. 2a) and the power correction diagrams from the
operators with $d=3$, $<\overline{q}q>$, $d=4$,
$m_{s}<\overline{q}q>$, $d=5$, $m_{0}^{2}<\overline{q}q>$  (Fig. 2b,
2c, 2d).
\begin{figure}
\vspace*{-1cm}
\begin{center}
\includegraphics[width=10cm]{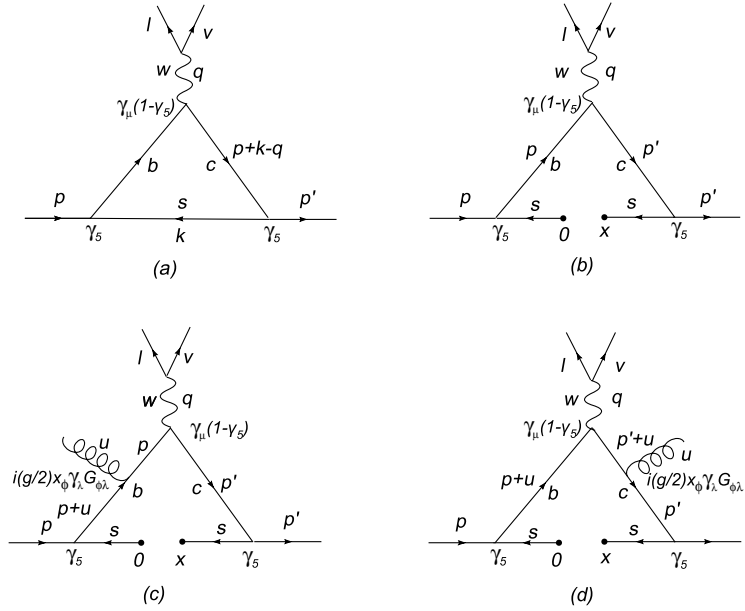}
\end{center}
\caption{Diagrams for bare-loop and power corrections (light quark
condensates) } \label{fig1}
\end{figure}
 In calculating the
bare-loop contribution, we first write the double dispersion
representation for the coefficients of corresponding Lorentz
structures appearing in the correlation function as:
\begin{equation}\label{10au}
\Pi_i^{per}=-\frac{1}{(2\pi)^2}\int ds\int
ds'\frac{\rho_{i}(s,s',q^2)}{(s-p^2)(s'-p'^2)}+\textrm{ subtraction
terms}.
\end{equation}
The spectral densities $\rho_{i}(s,s',q^2)$ can be calculated from
the usual Feynman integral (bare loop diagram in Fig. 2a) with the
help of Cutkosky rules, i.e., by replacing the quark propagators
with Dirac delta functions:
$\frac{1}{p^2-m^2}\rightarrow-2\pi\delta(p^2-m^2),$ which implies
that all quarks are real. After some straightforward calculations
for the  spectral densities corresponding to $P_{\mu}$ and $q_{\mu}$
we obtain:
\begin{eqnarray}\label{11au}
\rho_{1}(s,s',q^2)&=&N_{c}I_{0}(s,s',q^2)[2m_{b}m_{q}+2m_{c}m_{q}-4m_{q}^{2}
\nonumber\\
&-& 2A(\Delta+m_{b}^{2}-m_{q}^{2})-(A+B)u'-2B(\Delta'+m_{c}^{2}-m_{q}^{2})],\nonumber\\
\rho_{2}(s,s',q^2)&=&N_{c}I_{0}(s,s',q^2)[-2m_{b}m_{q}+2m_{c}m_{q}+2A(\Delta+m_{b}^{2}-m_{q}^{2})\nonumber\\
&+& (B-A)u'-2B(\Delta'+m_{c}^{2}-m_{q}^{2})],
\end{eqnarray}
where
\begin{eqnarray}\label{12}
I_{0}(s,s',q^2)&=&\frac{1}{4\lambda^{1/2}(s,s',q^2)},\nonumber\\
\lambda(s,s',q^2)&=&s^2+s'^2+q^4-2sq^2-2s'q^2-2ss',\nonumber \\
\Delta'&=&(s'-m_{c}^2 + m_{q}^2),\nonumber\\
\Delta&= &(s-m_{b}^2 + m_{q}^2),\nonumber\\
  A&=&\frac{1}{\lambda(s,s',q^2)}[2s'\Delta-\Delta'u],\nonumber\\
 B&=&\frac{1}{\lambda(s,s',q^2)}[2s\Delta'-\Delta u],\nonumber\\
u &=& s + s' - q^2,\nonumber\\
u'&=&2[m_{b}m_{c}-(m_{b}+m_{c})m_{q}+m_{q}^{2}].\nonumber\\
\end{eqnarray}
In Eq. (\ref{11au}) $N_{c}=3$ is the number of colors. The
integration region for the perturbative contribution in Eq.
(\ref{10au}) is determined from the condition that arguments of the
three $\delta$ functions must vanish simultaneously. The physical
region in $s$ and $s'$ plane is described by the following
inequalities:\\
\begin{equation}\label{13au}
-1\leq\frac{2ss'+(s+s'-q^2)(m_{b}^2-s-m_{q}^2)+(m_{q}^2-m_{c}^2)2s}{\lambda^{1/2}(m_{b}^2,s,m_{q}^2)\lambda^{1/2}(s,s',q^2)}\leq+1.
\end{equation}
From the above equation, it is easy to calculate the lower bound of
integration over $s'$ as a function of s. (i.e., $s'=f(s)$).

For the contribution of power corrections, i.e., the contributions
of operators with dimensions $d=3$, $4$ and $5$ (diagrams in Fig.
2b, 2c ,2d), we obtain the following results:
\begin{eqnarray}\label{14au}
\Pi^{non-per}_{1}&=&-\frac{1}{2r'r}<\overline{q}q>(m_{b}+m_{c})+\frac{m_{q}}{4}<\overline{q}q>
\left[\frac{m_{b}m_{c}+m_{c}^{2}}{r'^{2}r}+\frac{m_{b}^{2}+m_{b}m_{c}}{r'r^{2}}\right]\nonumber
\\
&+&
\frac{1}{2}<\overline{q}q>(-m_{q}^{2}+\frac{1}{2}m_{0}^{2})\left\{\frac{m_{b}m_{c}^{2}
+m_{c}^{3}}{r'^{3}r}+\frac{1}{2}\frac{(m_{c}+m_{b})(-q^{2}+m_{b}^{2}+m_{c}^{2})}{r'^{2}r^{2}}
\right. \nonumber \\
&+&
\left.\frac{1}{2}(m_{c}+m_{b})\left[\frac{1}{r'^{2}r}+\frac{1}{r'r^{2}}\right]+\frac{m_{b}^{3}+m_{b}^{2}m_{c}}{r'r^{3}}\right \} \nonumber \\
&+&<\overline{q}q>\frac{m_{0}^{2}}{48}\frac{(3m_{b}+m_{c})}{r'r^{2}}+<\overline{q}q>\frac{m_{0}^{2}}{48}\left\{\frac{-4(m_{b}+2m_{c})}{r'^{2}r}-
\frac{(2m_{b}+m_{c})}{r'r^{2}}\right. \nonumber \\
&+&\left.\frac{(-m_{b}^{3}+m_{b}^{2}m_{c}+m_{b}m_{c}^{2}-m_{c}^{3}+m_{b}q^{2}+m_{c}q^{2})}{r'^{2}r^{2}}\right\},
\nonumber \\
\Pi^{non-per}_{2}&=&\frac{1}{2r'r}<\overline{q}q>(m_{b}-m_{c})-\frac{m_{q}}{4}<\overline{q}q>
\left[\frac{m_{b}m_{c}-m_{c}^{2}}{r'^{2}r}+\frac{m_{b}^{2}-m_{b}m_{c}}{r'r^{2}}\right]
\nonumber \\
&+&
\frac{1}{2}<\overline{q}q>(-m_{q}^{2}+\frac{1}{2}m_{0}^{2})\left\{\frac{m_{b}m_{c}^{2}
-m_{c}^{3}}{r'^{3}r}+\frac{1}{2}\frac{(m_{c}-m_{b})(-q^{2}+m_{b}^{2}+m_{c}^{2})}{r'^{2}r^{2}}
\right. \nonumber \\
&+&
\left.\frac{1}{2}(m_{c}-m_{b})\left[\frac{1}{r'^{2}r}+\frac{1}{r'r^{2}}\right]-\frac{m_{b}^{3}-m_{b}^{2}m_{c}}{r'r^{3}}\right
\} \nonumber \\
&-&<\overline{q}q>\frac{m_{0}^{2}}{48}\frac{(3m_{b}-m_{c})}{r'r^{2}}+<\overline{q}q>\frac{m_{0}^{2}}{48}\left\{\frac{2(2m_{b}-m_{c})}{r'^{2}r}-
\frac{m_{c}}{r'r^{2}}\right. \nonumber \\
&+&\left.\frac{(m_{b}^{3}-3m_{b}^{2}m_{c}+3m_{b}m_{c}^{2}-m_{c}^{3}-m_{b}q^{2}+m_{c}q^{2})}{r'^{2}r^{2}}\right\},
\end{eqnarray}
where $r=p^2-m_{b}^2,r'=p'^2-m_{c}^2$.

The QCD sum rules for the form factors $f_{1}(q^{2})$ and
$f_{2}(q^{2})$ are obtained by equating the phenomenological and QCD
parts of the correlator  and applying double Borel transformations
with respect to the variables $p^2$ and $p'^2$ ($p^2\rightarrow
M_{1}^2,p'^2\rightarrow M_{2}^2$) in order to suppress the
contributions of higher states and continuum:
\begin{eqnarray}\label{15au}
f_{i}(q^2)&=&-\frac{(m_{b}+m_{q})
}{f_{B_{q}}m_{B_{q}}^2}\frac{(m_{c}+m_{q})}{f_{D_{q}}m^{2}_{D_{q}}}e^{m_{B_{q}}^2/M_{1}^2}e^{m_{D_{q}}^2/M_{2}^2}\nonumber\\
&&\bigg\{-\frac{1}{(2\pi)^2}\int_{(m_b+m_q)^2}^{s_0} ds
\int_{f(s)}^{s_0'}
ds'\rho_{i}(s,s',q^2)e^{-s/M_{1}^2}e^{-s'/M_{2}^2}\nonumber
\\&&+\emph{B}(M_{1}^{2})\emph{B}(M_{2}^{2})\Pi^{non-per}_{i}\bigg\},
\end{eqnarray}
where $i=1, 2$ and $\emph{B}(M_{1}^{2})\emph{B}(M_{2}^{2}) $ denotes
the double Borel transformation operator. In Eq. (\ref{15au}), in
order to subtract the contributions of the higher states and
continuum, the quark-hadron duality assumption is used, i.e., it is
assumed that
\begin{eqnarray}
\rho^{higher states}(s,s') = \rho^{OPE}(s,s') \theta(s-s_0)
\theta(s'-s'_0).
\end{eqnarray}
In calculations, the following rule for double Borel transformations
is also used:
\begin{equation}\label{16au}
\hat{B}\frac{1}{r^m}\frac{1}{r'^n}\rightarrow(-1)^{m+n}\frac{1}{\Gamma(m)}\frac{1}{\Gamma
(n)}e^{-m_{b}^2/M_{1}^{2}}e^{-m_{c}^2/M_{2}^{2}}\frac{1}{(M_{1}^{2})^{m-1}(M_{2}^{2})^{n-1}}.
\end{equation}
%
\section{Numerical analysis}
The sum rules expressions for the form factors $f_{1}(q^{2})$ and
$f_{2}(q^{2})$ show that the condensates, leptonic decay constants
of $B_{q}$ and $D_{q}$ mesons, continuum thresholds $s_{0}$ and
$s'_{0} $ and Borel parameters $M_{1}^2$ and $M_{2}^2$ are the main
input parameters.  In further numerical analysis, we choose the
value of the condensates at a fixed renormalization scale of about
$1$ GeV \cite{ioffe}:
$<\overline{d}d>=<\overline{u}u>=-(240\pm10~MeV)^3$,
$<\overline{s}s>=(0.8\pm0.2)<\overline{u}u>$ and
$m_{0}^2=0.8\pm0.2~GeV^2$.
  The experimental values for  the mass of the mesons,
 $m_{D_{s}}=1968.2\pm0.5~MeV$, $m_{D_{d}}(m_{D^{\pm}})=1869.3\pm0.4~MeV$,
$m_{D_{u}}(m_{D^{0}})=1864.5\pm0.4~MeV$
$m_{B_{s}}=5367.5\pm1.8~MeV$,
$m_{B_{d}}(m_{B^{0}})=5279.4\pm0.5~MeV$ and $
m_{B_{u}}(m_{B^{\pm}})=5279.0\pm0.5~MeV$ \cite{Yao} are used. For
the value of the leptonic decay constants and  quark masses, we use
the following values in two sets: In set 1, we use the results
obtained from two-point QCD sum rules analysis: $f_{B_{s}} = 209\pm
38~ MeV $ \cite{colangelo3}, $f_{D_{s}} =294\pm27 ~MeV $
\cite{colangelo2}, $f_{B_{d}} = f_{B_{u}} =140\pm10 ~ MeV $ \cite{T.
M.} and $f_{D_{d}} =f_{D_{u}}=170 \pm20~MeV $ \cite{T. M.}. The
quark masses are taken to be $ m_{c}(\mu=m_{c})=
 1.275\pm
 0.015~ GeV$, $m_{s}(1~ GeV)\simeq 142 ~MeV$ \cite{ming}, $m_{d}(1~ GeV)\simeq 5 ~MeV$, $m_{u}(1~ GeV)\simeq 1.5 ~MeV$ and $m_{b} =
(4.7\pm 0.1)~GeV$ \cite{ioffe}. In set 2, the recent  experimental
values   $f_{D_{s}} =274\pm13\pm7 ~MeV $ \cite{Artuso1}, $f_{D^{+}}
=222.6\pm16.7^{+2.8}_{-3.4} ~MeV $ \cite{Artuso2},  $f_{B^{+}}
=176^{+28+20}_{-23-19} ~MeV $ \cite{Yao} and lattice prediction for
$f_{B_{s}} = 206\pm 10~ MeV $ \cite {Rolf} are used. For heavy quark
masses $ m_{c}=1.25
 \pm
 0.09~ GeV$ and $m_{b}
 =4.7\pm0.07
~GeV$ \cite{Yao} and for  light quark masses the values at the scale
$\mu=1 GeV$ (the same as set 1) are considered.  The continuum
threshold parameters $s_{0}$ and $s_{0}' $ are also determined from
the two-point QCD sum rules: $s_{0} =(35\pm 2)~ GeV^2$
\cite{shifman} and $s_{0}' =6~ GeV^2 $ \cite{colangelo2}. The Borel
parameters $M_{1}^2$ and $M_{2}^2 $ are auxiliary quantities and
therefore the results of physical quantities should not depend on
them. In QCD sum rules method, OPE is truncated at some finite
order, leaving a residual dependence on the Borel parameters. For
this reason, working regions for the Borel parameters should be
chosen such that in these regions form factors are practically
independent of them. The working regions for the Borel parameters
$M_{1}^2 $ and $M_{2}^2$ can be determined by requiring that, on the
one side, the continuum contribution should be small, and on the
other side, the contribution of the operator with the highest
dimension should be small. As a result of the above-mentioned
requirements, the working regions are determined to be $ 10~ GeV^2 <
M_{1}^2 <22~ GeV^2 $ and $ 4~ GeV^2 <M_{2}^2 <10 ~GeV^2$.

In order to estimate the decay width of $B_{q} \rightarrow
D_{q}l\nu$ it is necessary to know the $q^2$ dependence of the form
factors $f_{1}(q^{2})$ and $f_{2}(q^{2})$ in the whole physical
region $ m_{l}^2 \leq q^2 \leq (m_{B_{q}} - m_{D_{q}})^2$. The $q^2
$ dependencies of the form factors can be calculated from QCD sum
rules (for details, see  \cite{ball1, ball2}). For extracting the
$q^2$ dependencies of the form factors from QCD sum rules, we should
consider a range of $ q^2$ where the correlation function can
reliably be calculated. For this purpose we have to stay
approximately $1~ GeV^2$ below the perturbative cut, i.e., up to
$q^2 =10 ~GeV^2$. In order to extend our results to the full
physical region, we look for parametrization of the form factors in
such a way that in the region $0 \leq q^2 \leq 10~ GeV^2$, this
parametrization coincides with the sum rules prediction. The
dependence of form factors $f_{1}(q^{2})$ and $f_{2}(q^{2})$  on
$q^2$ for set 1 are given in Figs. 3 and 4, respectively. Our
numerical calculations show that the best
parametrization of the form factors with respect to $q^2$ are as follows:\\
\begin{equation}\label{17au}
f_{i}(q^2)=\frac{f_{i}(0)}{1+ \alpha\hat{q}+ \beta\hat{q}^2+
\gamma\hat{q}^3+ \lambda\hat{q}^4},
\end{equation}
where $\hat{q}=q^2/m_{B_{q}}^2$. The values of the parameters
 $f_{i}(0), ~\alpha,~ \beta,~ \gamma$ and $\lambda$ for set1 are
given in the Table 1.

Now, we are going to calculate the total decay width for these
transitions. The differential decay  width is  as follows:
\begin{eqnarray}\label{29au}
\frac{d\Gamma}{dq^2}&=&\frac{1}{192\pi^{3}m_{B_{q}}^{3}}
G_{F}^2|V_{cb}|^2\lambda^{1/2}(m_{B_{q}}^{2},m_{D_{q}}^{2},q^{2})\left(\frac{q^{2}-m_{\ell}^{2}}{q^{2}}\right)^{2}
\nonumber \\
&\times&\left\{-\frac{1}{2}(2q^{2}+m_{\ell}^{2})\left[|f_{1}(q^{2})|^{2}(2m_{B_{q}}^{2}+2m_{D_{q}}^{2}-q^{2})
\right.\right. \nonumber \\
&+&\left.\left.2(m_{B_{q}}^{2}-m_{D_{q}}^{2})Re[f_{1}(q^{2})f_{2}^{*}(q^{2})]+|f_{2}(q^{2})|^{2}q^{2}\right]\right.\nonumber
\\
&+&\left.\frac{(q^{2}+m_{\ell}^{2})}{q^{2}}\left[|f_{1}(q^{2})|^{2}(m_{B_{q}}^{2}-m_{D_{q}}^{2})^{2}
\right.\right. \nonumber \\
&+&\left.\left.2(m_{B_{q}}^{2}-m_{D_{q}}^{2})q^{2}Re[f_{1}(q^{2})f_{2}^{*}(q^{2})]+|f_{2}(q^{2})|^{2}q^{4}\right]
\right\}.
\end{eqnarray}
\begin{table}[b]
\centering
\begin{tabular}{|c|c|c|c|c|c|} \hline
& f(0)  & $\alpha$ & $\beta$& $\gamma$& $\lambda$\\\cline{1-6}
 $f_{1}(B_{s}\rightarrow D_{s}(1968)\ell\nu)$ & 0.24 & -1.57 & 1.66& -10.43& 19.06\\\cline{1-6}
 $f_{2}(B_{s}\rightarrow D_{s}(1968)\ell\nu)$ & -0.13  & -1.69 & 0.11& 1.50& -4.65\\\cline{1-6}
 $f_{1}(B_{u}\rightarrow D_{u}(1864)\ell\nu)$ & 0.52  & -1.49 & 0.02 & 0.93& -3.76\\\cline{1-6}
 $f_{2}(B_{u}\rightarrow D_{u}(1864)\ell\nu)$ & -0.29  & -1.69 & 0.21& 0.90& -3.38\\\cline{1-6}
 $f_{1}(B_{d}\rightarrow D_{d}(1869)\ell\nu)$ & 0.52  & -1.49 & 0.05 & 0.77& -3.47\\\cline{1-6}
 $f_{2}(B_{d}\rightarrow D_{d}(1869)\ell\nu)$ & -0.29  & -1.69 & 0.16& 1.13& -3.70\\\cline{1-6}
 \end{tabular}
 \vspace{0.8cm}
\caption{Parameters appearing in the form factors of the
$B_{q}\rightarrow D_{q}\ell\nu$}decays in a four-parameter fit for
$M_{1}^2=15~GeV^2$, $M_{2}^2=6~GeV^2$ and set1. \label{tab:1}
\end{table}

Next step is to calculate  the value of the branching ratio for
these decays. Taking into account the $q^2$ dependencies of the form
factors and performing integration over $q^2$ in Eq. (\ref{29au}) in
the interval $m_{l}^2\leq q^2\leq(m_{B_{q}}-m_{D_{q}})^2$ and using
the total life-time $\tau_{B_{s}}=1.46\times10^{-12}s$
\cite{eidelman}, $\tau_{B_{d}}=1.64\times10^{-12}s$,
$\tau_{B_{u}}=1.53\times10^{-12}s$ \cite{Yao} and $\mid
V_{cb}\mid=0.0416\pm0.0006$ \cite {Ceccucci},  the following results
of the branching ratios for set 1 are obtained.
\begin{eqnarray}\label{}
\mbox{\textbf{B}}(B_{s}\rightarrow D_{s}\ell \nu)=(2.8-3.5) \times
10^{-2},\nonumber \\
\mbox{\textbf{B}}(B_{d}\rightarrow D_{d}\ell \nu)=(1.8-2.4) \times
10^{-2},\nonumber \\
\mbox{\textbf{B}}(B_{u}\rightarrow D_{u}\ell \nu)=(1.6-2.2) \times
10^{-2}.
\end{eqnarray}
 The result for $B_{s}\rightarrow D_{s}\ell\nu$ shows that this transition   can also be easily detected
 at LHC in the near future. The measurements of this channel and
 comparison of their results with that of  the phenomenological methods like
 QCD sum rules could give useful information about the structure of
 the $D_{s}$ meson.

At the end of this section, we would like to compare the present
work results of   the form factors and their limits at heavy quark
effective theory (HQET) (for details see  \cite{aliev4}) for two
sets with the predictions of the lattice QCD \cite{Hashimoto,
Divitiis2} at zero recoil limit for $B\rightarrow D\ell\nu$. For
this aim, we introduce the notations used in \cite{Hashimoto,
Divitiis2} equivalent to Eq. (\ref{3au})
\begin{equation}\label{complatt}
<D\mid\overline{c}\gamma_{\mu} b\mid
B>=\sqrt{m_{B}m_{D}}\bigg[h_{+}(v+v')_{\mu}+h_{-}(v+v')_{\mu}\bigg],
\end{equation}
where $h_{+}$ and  $h_{-}$ are the transition form factors and $v$
and $v'$ are the four velocities of the initial and final meson
states. The relations between our form factors with the $h_{+}$ and
$h_{-}$  are given as:
\begin{eqnarray}\label{}
f_{1}=\frac{(m_{B}+m_{D})h_{+}-(m_{B}-m_{D})h_{-}}{2\sqrt{m_{B}m_{D}}},\nonumber \\
f_{2}=\frac{(m_{B}+m_{D})h_{-}-(m_{B}-m_{D})h_{+}}{2\sqrt{m_{B}m_{D}}}.
\end{eqnarray}
In order to perform the heavy quark mass limit, we define the
multiplication of the $v$ and $v'$ as
 \begin{equation}\label{wq}
 w=vv'=\frac{m_{B}^2+m_{D}^2-q^2}{2m_{B}m_{D}}.
 \end{equation}
 At zero recoil limit, $w=1$ and from Eq. (\ref{wq}) it is correspond
 to $q^2\simeq 11~GeV^2$ which lies in the interval $m_{l}^2\leq
 q^2\leq(m_{B}-m_{D})^2$. Table 2 shows a comparison of the form factors
 and their HQET limits in the present work and the lattice QCD
predictions at HQET  and zero recoil ($w=1$) limits  in the present
study notations.
\begin{table}[h]
\centering
\begin{tabular}{|c|c|c|} \hline
& $f_{1} $ & $f_{2}$
\\\cline{1-3}
Present study-set1 & $1.29 \pm0.15$ & $-0.83\pm0.10$   \\\cline{1-3}
Present study (HQET )-set1& $1.24 \pm0.12$ & $-0.68\pm0.08$
\\\cline{1-3}Present study -set2& $1.10 \pm0.14$ & $-0.72\pm0.09$   \\\cline{1-3}
Present study (HQET )-set2& $1.06 \pm0.10$ & $-0.58\pm0.06$
\\\cline{1-3} Lattice QCD (HQET )\cite {Hashimoto} & $1.19\pm0.01$ &
$-0.68\pm0.05$
\\\cline{1-3}
 Lattice QCD (HQET )\cite {Divitiis2} & $1.16\pm0.03$ & $-0.56\pm0.05$ \\\cline{1-3}
\end{tabular}
\vspace{0.8cm} \caption{ Comparison of the form factors in the
present work, their HQET limits  and lattice QCD predictions  at
HQET and zero recoil ($w=1$) limits  in the present study
notations.} \label{tab:3}
\end{table}
 From  this Table, it is clear that there is a good consistency among the models especially when we consider the errors.
  Moreover, a comparison of  our
results for the branching ratio of the $B_{q}\rightarrow
D_{q}\ell\nu$ with the predictions of the CQM model \cite{zhao} and
the experiment \cite{Yao} are also given in Table 3. Considering the
uncertainties and intervals, this Table also shows a good agreement
among the phenomenological approaches and the experiment.
Furthermore, this Table  indicates that the value of the branching
ratio increases both in the present work and the experiment by
increasing the mass of the q quark. The intervals and uncertainties
for values in the present study are related to the uncertainties in
the values of the input parameters as well as different lepton types
$(e, \mu, \tau)$. Our results for set 1 and set2 show that the value
of the branching ratio is sensitive to the uncertainties in the
value of the leptonic decay constants as well as the heavy quark
masses. The existing uncertainties in light quark masses for $q= u$
and $ d$ cases don't change the results but for $q=s$ case, we see a
variation about $3^{0}/_{0}$ in the value of the branching ratio.

 In conclusion, the semileptonic $B_{q}\rightarrow D_{q}\ell\nu$
 decays were investigated in QCD sum rules method. The $q^2$ dependencies of
the transition form factors were evaluated. Using the expressions
for the related form factors, the total decay width   and the
branching ratio for these decays  have been estimated. The results
enhance the possibility of  observation of the $B_{s}\rightarrow
D_{s}\ell\nu$ at LHC in the near future. Finally, the comparison of
our results with that of the other phenomenological approach,
lattice QCD and experiment was presented.

\begin{table}[h]
\centering
\begin{tabular}{|c|c|c|c|} \hline
& $(B_{s}\rightarrow D_{s}\ell \nu) $ & $(B_{d}\rightarrow
D_{d}\ell \nu)$ & $(B_{u}\rightarrow D_{u}\ell \nu) $
\\\cline{1-4}
Present study-set1 & $(2.8-3.5) \times 10^{-2}$ & $(1.8-2.4) \times
10^{-2}$ & $(1.6-2.2) \times 10^{-2}$\\\cline{1-4}Present study-set2
& $(3.0-3.8) \times 10^{-2}$ & $(1.5-2.2) \times 10^{-2}$ &
$(1.3-2.0) \times 10^{-2}$\\\cline{1-4} CQM model & $(2.73-3.0)
\times 10^{-2}$ & $(2.2-3.0) \times 10^{-2}$ & $(2.2-3.0) \times
10^{-2}$\\\cline{1-4} Experiment & - & $(2.15 \pm 0.22) \times
10^{-2}$ & $(2.12 \pm 0.2) \times 10^{-2}$\\\cline{1-4}
\end{tabular}
\vspace{0.8cm} \caption{ Comparison of the branching ratios for
$B_{q}\rightarrow D_{q}\ell\nu$} decays in QCD sum rules approach,
the CQM model \cite{zhao} and the experiment \cite{Yao}.
\label{tab:3}
\end{table}

\section{Acknowledgment}
  The author would like to thank T. M. Aliev and A. Ozpineci for their useful discussions and TUBITAK,
  Turkish scientific and research council, for their partially
  support.

\clearpage
\begin{figure}
\vspace*{-1cm}
\begin{center}
\includegraphics[width=10cm]{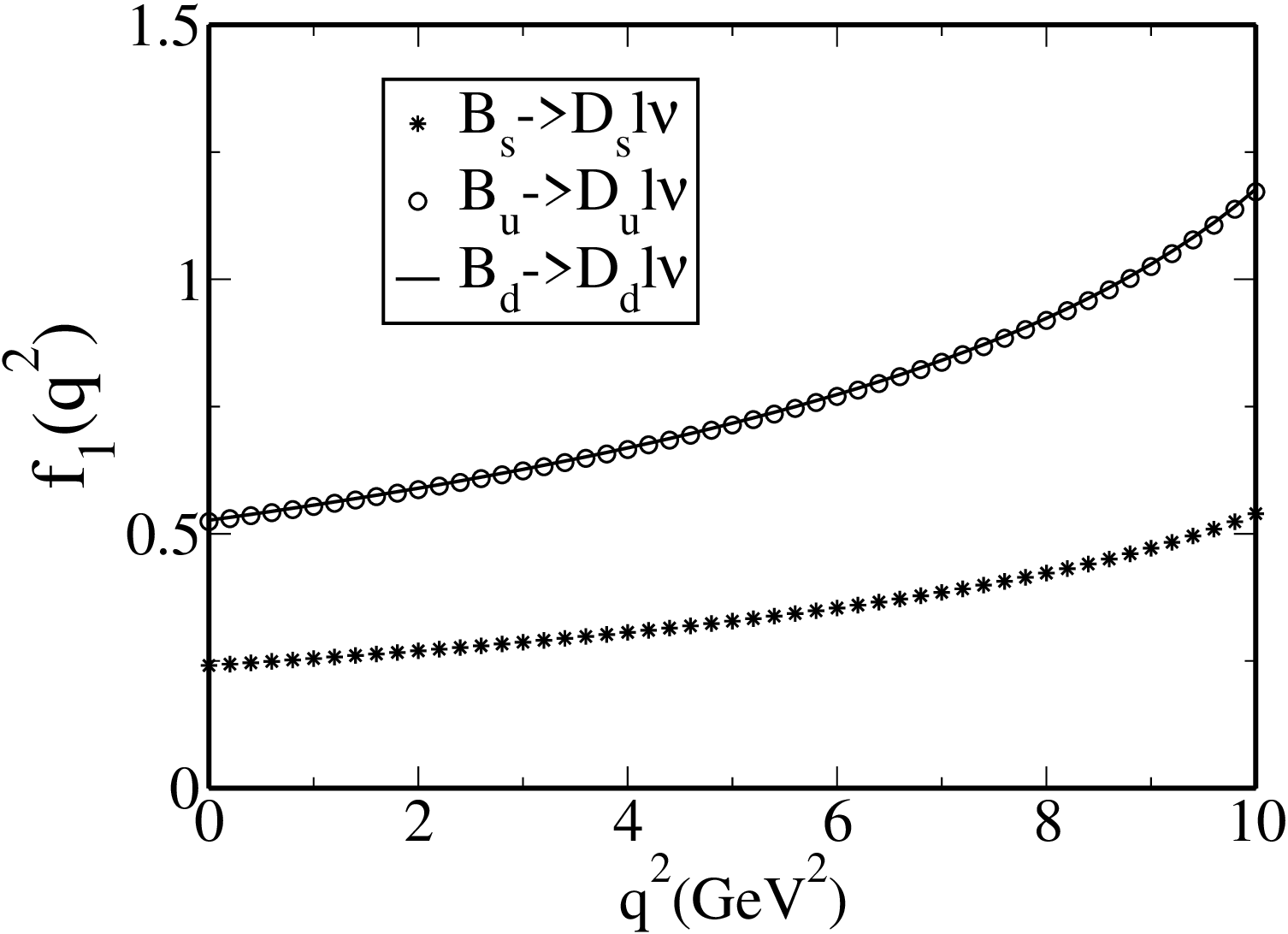}
\end{center}
\caption{The dependence of $f_{1}$ on
 $q^2$ at $M_{1}^2=17~GeV^2$, $M_{2}^2=6~GeV^2$, $s_{0}=35~GeV^2$, $s_{0}'=6~GeV^2$ and set1. } \label{fig3}
\end{figure}
\begin{figure}
\vspace*{-1cm}
\begin{center}
\includegraphics[width=10cm]{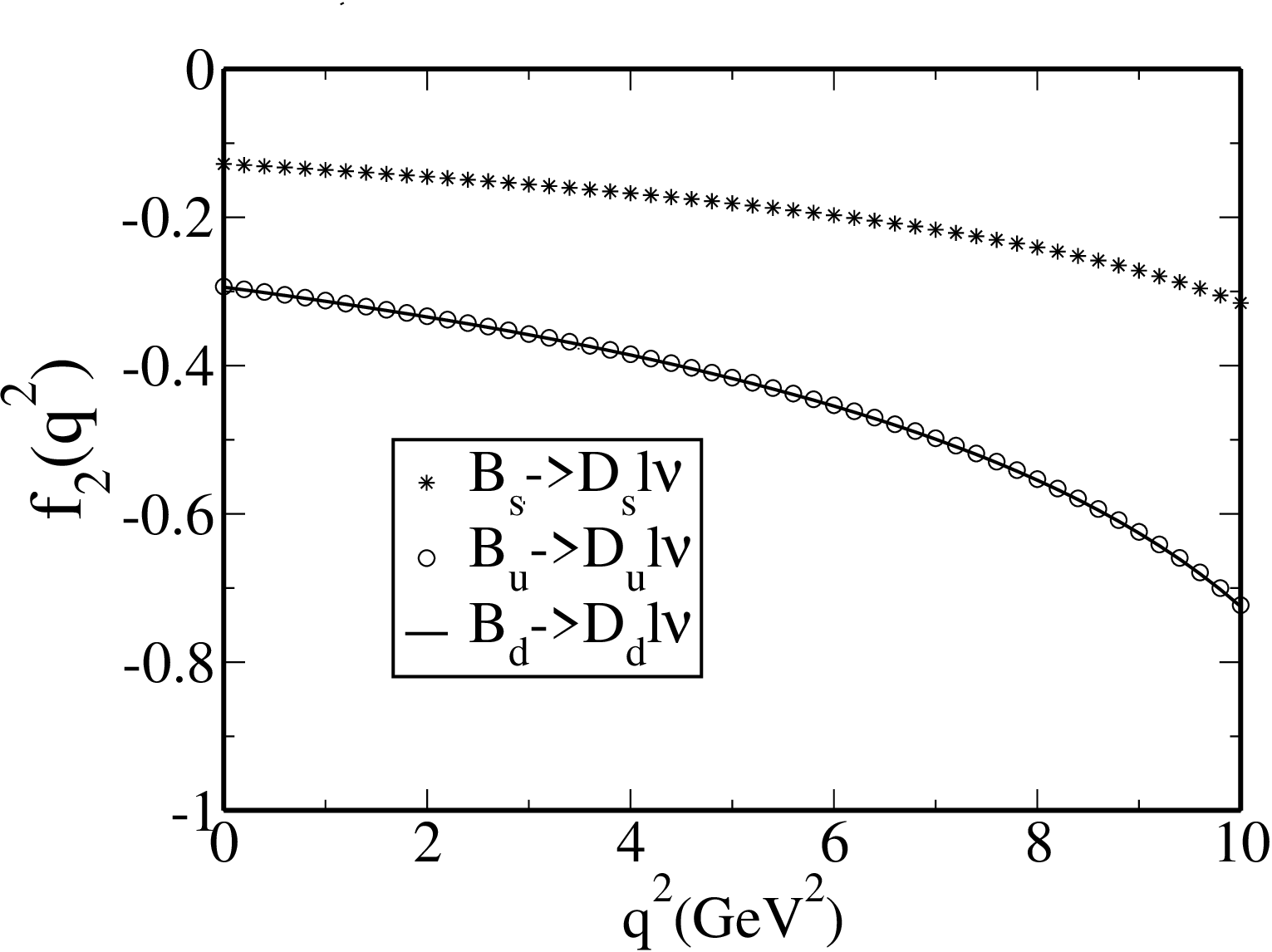}
\end{center}
\caption{The dependence of $f_{2}$ on
 $q^2$ at $M_{1}^2=17~GeV^2$, $M_{2}^2=6~GeV^2$, $s_{0}=35~GeV^2$, $s_{0}'=6~GeV^2$ and set1. } \label{fig1}
\end{figure}
\end{document}